\documentclass[10pt,twoside]{article}
\usepackage{amsmath}
\usepackage{amsfonts}
\usepackage{amssymb,bbm}
\usepackage{makeidx}
\def\bea{\begin{eqnarray}}
\def\eea{\end{eqnarray}}

\newcommand{\SR}{\left( \pm \right)}
\newcommand{\SP}{\left( + \right)}
\newcommand{\SM}{\left( - \right)}
\author{Rudranil Basu \footnote{rudranil@bose.res.in}\\ Samir K Paul \footnote{smr@bose.res.in}\\ S.N. Bose National Centre for Basic Sciences, Sector-III,\\ Block - JD, Salt Lake, Kolkata - 700 098.}
\title{~~~~Consistent 3D Quantum gravity on Lens Spaces}
\begin{document}
\maketitle
\begin{abstract}
We study non-perturbative quantization of 
the first order formulation of 3d gravity  with positive cosmological constant (de Sitter space being the prototype vacuum solution, whose Euclideanization 
gives the three sphere) on the background topology of lens space, which is a three spheres modulo a discrete group. Instead of the strategy followed by a recent work \cite{Castro:2011xb}, which compares results in the second and first order formulations of gravity, we concentrate on the later solely. We note, as a striking feature, that the quantization which relies heavily on the axiomatic of topological quantum field theory (TQFT), can only be consistently carried by augmenting the conventional theory by an additional topological term coupled through a dimensionless parameter. More importantly the introduction of this additional parameter renders the theory finite.
\end{abstract}
\section{Introduction}
Most of the non-trivial results in 3d gravity including the famous BTZ black hole solution is known for the negative cosmological constant sector. Also there is a definite trace  of AdS/CFT correspondence when the space-time is asymptotically AdS. On the other hand study of 3d gravity with positive cosmological constant has generated considerable interest only recently \cite{Castro:2011xb}. This involves evaluation of 1 loop partition function in the metric formulation in order to find the de Sitter vacuum, namely the Hartle Hawking state. They showed for the first time the equivalence of Chern Simons framework of gravity with Einstein theory up to 1-loop level in the quantum regime. 
 In addition to that topologically massive gravity (TMG), which unlike pure gravity consists of propagating modes, has been thoroughly studied in \cite{Castro:2011ke}. The main question these studies aim to address is how one can make sense of 3d de Sitter quantum gravity, through the vacuum state. Surprisingly enough, the pure topological gravity theory fails to give any satisfactory answer to it in the sense that the partition function (both in one loop and nonperturbative computations) tend to diverge unregularizably when one considers sum over the infinitely large class of lens spaces (a typical solution of 3D de Sitter gravity, corresponding to saddle points in the path integral); whereas the answer for TMG containing local degrees of freedom is in the affirmative. The latter is tame under sum over topologies.

The pure gravity and TMG calculations have been considered  in the Euclidean signature with the motivation that Euclideanized de Sitter gravity is `thermal'. This has been made precise in terms of the Euclidean de Sitter geometry in \cite{Castro:2011xb}. Moreover, in the Einstein-Hilbert theory path integral is sensible in the Euclidean picture. 
On the other hand if one prefers to study the theory in first order formulation, in the Chern-Simons (CS) framework, Euclideanization is not an obvious idea that one should come across. This is because CS theory is manifestly topological and doesn't rely on background metric as long as perturbative analysis remains not as the primary goal. 
 But once one tries to make contact with metric formulation through $ \langle e _{\mu}, e_{\nu} \rangle = g _{\mu \nu}$, Euclideanization can be viewed from the choice in the internal metric on the frame bundle (of vielbeins), and hence the structure group. This change reflects upon the choice of gauge group of the CS theory. Gauge group changes from non-compact $SO(3,1)$ to compact $SU(2) \times SU(2) $, thus making the problem tractable from gauge theory perspective. The action then becomes difference of two $SU(2)$ CS theories.

This is the motivation for our purpose to look at Euclideanized version. In this case we don't need a Wick rotation in space time 
and our partition functions keeps the formal expression
$$ Z = \int DA \exp \left(i \frac{k}{4 \pi}\int \mathrm{tr}\left(A \wedge dA + \frac{2}{3} A^3\right)\right).$$
where `tr' stands for the metric over $\mathfrak{su}(2)$ .
We would see that this form of the path integral will help us in the end so that the trouble of working with imaginary coupling of CS won't also come in our way \footnote{In another important work Witten \cite{Witten:2010cx} recently pointed out how quantization of CS theory with complex coupling can be carried out by suitably deforming the functional integral contour. However for this case one still has to study the possibility of associating a finite dimensional Hilbert space of CS theory on a compact Riemann surface, which we need for quantization here.}. Since we would be confined in the first order regime, our concern about the background appears only through its possible topologies. The choice of topology is however motivated strongly from the fact that Euclideanized de Sitter space can be identified with $S^3$, through its metric and topology. We choose it to be of the form $S^3 / \Gamma$, or lens spaces to be precise. $\Gamma$ is a suitable discrete group with known action on $S^3$ as in \cite{Castro:2011xb}. Of course feasible solutions are always locally dS.

Now at this point it may seem that we are free to choose any of the standard quantization techniques for this theory. This may involve directly evaluating the partition function or taking recourse to geometric quantization \cite{Axelrod:1989xt}. As is well known that the former is well suited for perturbative calculations and computation of determinant of the elliptic operator that arises is well understood in terms of the analytic torsion even for non-compact gauge groups \cite{BarNatan:1991rn}. But once we are interested in nonperturbative results we must investigate the gauge moduli space of solutions, upon which a suitable canonical quantization may be carried out. However, on the given topology of lens space the solution space modulo the gauge transformations give only a collection of finite points, which certainly isn't a symplectic manifold. We therefore use standard surgery and gluing prescription for the construction of the space and using axioms of TQFT find the partition function as \cite{Jeffrey:1992tk} \footnote{Choosing framing of surgery suitably.}
\bea \label{part1}
Z = \langle \psi | U | \psi \rangle .\eea
Here $ | \psi \rangle \in {\cal H}_{T^2}$ is a state of quantized CS theory on the boundary of a solid torus, gluing two of which we construct a lens space. $U$ is an element of the $T^2$ mapping class group, specifying which gives us a class of lens spaces. This is where `conventional wisdom' of viewing first order gravity as difference of two $SU(2)$ CS theories fails. This failure becomes manifest when one looks at the CS levels $\pm \dfrac{l}{8G}$ ($l$ being inverse of the root of the cosmological constant and $G$ the Newton's constant). 

But we see that in the famous work of Witten \cite{Witten:1988hc}, a plausible approach of viewing first order gravity theory as a difference of two CS with {\it unequal levels} were presented. Ab initio this action does not have a metric interpretation. Nevertheless it gives same equations of motion as that of ordinary CS gravity, which are  equivalent to Einstein's equation for the class of invertible vierbeins. The crux is that as one solves the torsionless condition (half of the equations of the motion) and substitutes in the action, it becomes metric TMG and gains local excitations. This is very much unlike the case of CS gravity with equal and opposite couplings. However within the arena of first order gravity alone one could still get back metric interpretation through construction of a dual CFT , especially in the negative cosmological constant sector. First step towards this exciting result was taken in \cite{231928} in the metric framework. In a more recent work (although for negative cosmological constant) \cite{Witten:2007kt} this approach has been proved to work well in terms of holomorphically factorizable dual CFTs for CS gravity wih unequal couplings. Chiral and anti-chiral central charges are presented there in terms of the CS couplings (also see \cite{Basu:2011qy}) and quantum BTZ black holes are studied. The resulting CFT has been shown to be reach in content in reference to the monster group.

The same theory of gravity as two $SU(2)$ CS with unequal couplings has been studied in \cite{Bonzom:2008tq} where geometrical observables like area and length are quantized for their spectra. This illuminates that one can study quantum theories involving metric even without starting with a theory of metric variable. The new parameter enters the spectra in a way that it makes the spetra physically meaningful.

The problem with equal and opposite couplings of is that the CS part corresponding to the negative level is ill-defined and cannot be quantized on $T^2$ \cite{Basu:2009dy}. We need to extend the theory in a way described in \cite{Witten:1988hc,Basu:2009dy,Bonzom:2008tq,Witten:2007kt} so that the couplings of the CS theories can be tuned to be positive. This is a necessary condition since dim(${\cal H}_{T^2}$) equals the product of shifted CS couplings. When both the couplings are positive integers we get a situation which we regard as {\it consistent quantization}. At the same time it is worthwhile to mention that such an extension does not alter the equations of motion. Hence the gravitational interpretation of the theory remains intact.

Furthermore due to this extension (through introduction of a new dimensionless parameter) we get a finite answer for the partition function, as opposed to \cite{Castro:2011xb}. We exhibit the finiteness explicitly at a certain limit of this new parameter. This is certainly an improvement towards finding an answer about how meaningful 3d de Sitter quantum gravity is.
\section{The Extended Theory}
Functional of the $SU(2)$ vielbein and connection, the conventional Euclidean theory describing first order 3d gravity with positive cosmological constant $ \Lambda = \dfrac{1}{l^2}$ is (in the units where $16\pi G =1 =c$)
\bea \label{action1}
S[e,\omega] = 2\int \left( e^I \wedge \left( 2 d \omega _I + \epsilon_{IJK} \omega ^J \omega ^K \right) + \frac{1}{3 l^2} \epsilon _{IJK} e ^I \wedge e^J \wedge e^K \right)
\eea
In terms of $SU(2)$ CS connections $A ^{\SR} = \omega \pm e/l$ this action reads
$$ S = l \left( I[A^{\SP}] - I[A^{\SM}]\right)$$
with 
$$I[A] = \int \left( A^{ I} \wedge d A _I + \frac{1}{3} \epsilon _{IJK} A^{ I} \wedge A^{J} \wedge 
A^{K} \right)$$ as the integral of the CS form. The variational problem is perfectly well defined on the topology of the lens space $ S^3 / {\Gamma}$. Equations of motion are the well known ones:
$$\mathrm{flat ~CS ~connections}~~~2dA^{\SR}_I + \epsilon_{IJK} A^{\SR J} \wedge A^{\SR K} =0$$ or in terms of variables pertaining to gravity:
\begin{subequations} \label{eom}
\bea
&&\mathrm{torsionless ~condition} ~~~~ d e^I + \epsilon ^{IJK} e_J \wedge \omega _K =0 \label{eoma}~~\mbox{and} \\
&&\mathrm{curvature ~equation} ~~~~ 2 d \omega ^I + \epsilon ^{IJK} \omega_J \wedge \omega _K = - \frac{1}{l^2} \epsilon ^{IJK} e_J \wedge e _K \label{eomb}
\eea
\end{subequations}

Now the observation that the action 
\bea \label{action2}
\tilde{S}[e, \omega]&=&2l\int \left(\omega^I \wedge d \omega _I + \frac{1}{l^2} e^I \wedge d e_I + \frac{1}{3} \epsilon _{IJK} \omega^I \wedge \omega^J \wedge\omega^K + \frac{1}{l^2}\epsilon _{IJK} \omega^I \wedge e^J \wedge e^K \right) \nonumber\\
&=& l \left( I[A^{\SP}] + I[A^{\SM}] \right)
\eea
on a closed manifold also gives the same equations of motion \eqref{eom} motivates one to linearly combine \eqref{action2} to \eqref{action1}. In terms of the CS variables, one therefore constructs the action with introduction of a new parameter $ \gamma$:
\bea \label{action3}
\tilde{I} [A^{\SP},A^{\SM}] &=& S+ \frac{1}{\gamma} \tilde{S} \nonumber\\
&=&\frac{k_{\SP}}{2\pi} I[A^{\SP}] + \frac{k_{\SM}}{2\pi} I[A^{\SM}] 
\eea
where $k_{\SR} =\dfrac{l\left( 1/ \gamma \pm 1\right)}{8 G} $. Here we have restored $G$ so that Einstein's equation is satisfied .

It now calls for a short discussion for interpreting \eqref{action3}. 
Most interestingly it gives the same equations of motion \eqref{eom} (for manifolds without boundary), independent of the couplings $k_{(\pm)}$. This feature was first noted in the celebrated paper \cite{Witten:1988hc}. When \eqref{eoma} is solved for $ \omega$ and substituted in the \eqref{eomb}, one exactly gets the Einstein equation of the metric theory for the invertible class of vierbeins from (3). More detailed description about this theory and its relationship with Einstein-Hilbert theory and TMG is available in \cite{Grumiller:2008pr},\cite{Basu:2009dy}.

Although the solution space for this extended theory remains same, we find that the phase space structures are different. The presymplectic structure of the theory given in terms of two arbitrary vector fileds tangential to the space of solutions is
\bea \label{symp}
\Omega \left( \delta _1 , \delta _2\right) = \frac{k _{\SP}}{\pi} \int _{\Sigma} \delta _1 A ^{\SP} \wedge \delta _2 A ^{\SP} + \frac{k _{\SM}}{\pi} \int _{\Sigma} \delta _1 A ^{\SM} \wedge \delta _2 A ^{\SM}.
\eea
$\Sigma$ is a suitable Cauchy foliation of the base manifold. It is clear that the situation $k_{\SM} \rightarrow 0$ as $ \gamma \rightarrow 1$, is comparable to the `chiral point' of the theory in the AdS case, which has a well understood dual CFT. At this point the pre-symplectic structure automatically becomes degenerate in the $ \delta A^{\SM}$ directions (leaving apart its original gauge degeneracy). This degeneracy is evident if one considers the equal Euclidean time Poisson brackets:
\bea \label{PB2}
\{\omega^{ I}_i (x,\tau),e^{ J}_j (y,\tau)\}&=& 4\pi G \frac{\gamma ^2}{\gamma ^2 -1} \varepsilon _{ij} \delta ^{IJ} 
\delta ^2 \left(x,y \right)\nonumber\\
\{\omega^{ I}_i (x,\tau),\omega^{ J}_j (y,\tau)\} &=& -4\pi G \frac{\gamma /l}{\gamma ^2 -1} \varepsilon _{ij} \delta ^{IJ} 
\delta ^2 \left(x,y \right) \\
\{e^{ I}_i (x,\tau),e^{ J}_j (y,\tau)\} &=& -4\pi G \frac{\gamma l}{\gamma ^2 -1} \varepsilon _{ij} \delta ^{IJ} \delta ^2 
\left(x,y \right) \nonumber ;
\eea
$\delta ^{IJ}$ is the $ \mathfrak{su} (2)$ metric.
\section{Problems with canonical quantization on lens space}
Since we are interested in the nonperturbative evaluation of the partition function, the information about Lens space that suffices is its algebraic topology. This is given by \footnote{Role of $q ({\mathrm{mod}}p)$ coprime to $p$ comes through the action ${\mathbb Z}_p : S^3 \rightarrow S^3$. This is most easily viewed by considering $S^3$ as unit sphere in $ \mathbb{C}^2$ and specifying the ${\mathbb Z}_p$ action as $\left( z_1, z_2\right) \mapsto \left({\mathrm e} ^{2 \pi i /p}z_1 , {\mathrm e} ^ {2 \pi i q/p} z_2\right)$.} $L(p,q) = S^3 / {\mathbb Z}_p$. The physical phase space of this theory containing only flat connections, is given by $\left( \hom : \pi _1 \left(L(p,q)\right) \rightarrow SU(2) \right)/\sim$, (moduli space of flat $SU(2)$ connections modulo gauge transformations) where $\sim$ denotes gauge equivalence classes. For lens space $L(p,q)$, the fundamental group is isomorphic to ${\mathbb Z}_p$, which is freely generated by a single generator, say $ \alpha $; ie the group consists of the elements $\{ \alpha ^n | n=0,\dots ,p-1\}$. The homomorphisms to $SU(2)$, which we denote by $h$ must satisfy $h[ \alpha ^p] = \left(h[ \alpha] \right)^p = \mathbbm{1}$. In the defining representation (using the freedom of group conjugation) of $SU(2)$, this gives 
$$ h[ \alpha] = \mathrm{e} ^{2\pi i \sigma _3 /p}.$$
Hence the moduli space consists of only $p$ distinct points and therefore can in no way be a symplectic manifold. In physical terms these points represent holonomies of the $p$ disjoint non contractible loops around the $p$ marked points on $L(p,q)$.

In this connection we wish to emphasize that the configuration corresponding to $n=0$ above, is unique to first order gravity only. It represents the holonomy of the connection $A^{\SR} =0 $ or its gauge equivalent class. This means that we are taking the $e=0=\omega$ solution in our phase space. These configurations do not give rise to any physically meaningful metric, as elucidated in \cite{Witten:2007kt}. But while doing non-perturbative quantization of first order theory we must include them in the phase space.
\section{Appropriate quantization}
\subsection{$ {\mathcal{H}} _{T^2}$}
That we have seen direct attempts to quantize the theory on $L(p,q)$ fails, we should resort to indirect means as exemplified in \eqref{part1}. In this respect we construct $L(p,q)$ by gluing two solid tori through their boundaries using an element of the mapping class group
\bea \label{U} U= \begin{pmatrix} q & b \\ p & d
\end{pmatrix} \in SL(2, \mathbb{Z}).\eea
The quantization strategy \cite{Jeffrey:1992tk} as outlined in the introduction requires associating two quantum Hilbert spaces of the CS theory with the boundary of the solid tori. We therefore have to find $ {\cal H}_{T^2}$. Although this can be found in various places, for example in \cite{Axelrod:1989xt, Basu:2009dy, Atiyah:1990dn, Elitzur:1989nr}, for completeness we would like to give a simple and short description of it.

Since we are quantizing CS theory on $T^2$ (the third dimension may be taken as $ \mathbb{R}$, the whole 3 manifold being viewed as a trivial line bundle over $ T^2$), we have as the starting point, the moduli space : $\left( \hom : \pi _1 (T^2) \rightarrow SU(2) \right)/\sim$.

Now $ \pi _1 (T^2) = \mathbb{Z} \oplus \mathbb{Z}$ and is a freely generated abelian group with two generators $ \alpha ,\beta$ having the relation $ \alpha \beta \alpha ^{-1} \beta ^{-1} = 1$. Taking privilege of the group conjugacy as before we take the 2 dimensional representation of the homomorphism maps as:

\bea \label{hom}
h [ \alpha] = \mathrm{e}^{i \sigma _3 \theta} ~~~~ h [ \beta] = \mathrm{e}^{i \sigma _3 \phi} ~~~~ \theta, \phi \in [-\pi, \pi] .
\eea
This endows the two dimensional moduli space $ \mathcal{M}$ with the topology of  $ T^2$ (parameterized by $ \theta , \phi$). Note that this simple construction of $\cal M$ is motivated from the rigorous point of viewing it as $ {\cal M} = T \times T / W $, where $T$ is the torus of maximal dimension (for $SU(2)$ which is 1 and $T=S^1$) and $W$ is the Weyl group with $Ad$ action on the group. Our strategy will be to first quantize $ T \times T$ and then take Weyl invariant `parallel' sections of the line bundle on it. 

The `pushed down' symplectic structure on $ \mathcal{M}$ is
\bea \label{symp2}
\omega = \frac{k}{2\pi} d \theta \wedge d\phi .\nonumber
\eea
An appeal to Weil's integrality criterion 
\bea \label{weil} \int _{ \cal M} \frac{ \omega}{2\pi} \in \mathbb{Z}\eea now assures that $k$ must be an integer. At the stage of prequantization a prequantum line bundle is chosen over $ \cal M$ and before choosing the polarization for this line bundle we pick a complex structure $ \tau$ for $ \cal M$ (induced by that on the surface of the solid torus). This gives us the holomorphic coordinate: $ z = \frac{1}{\pi}( \theta + \tau \phi)$ on $ \cal M$. We re-express $$\omega = \frac{i k \pi}{4 \tau _2} d z \wedge d \bar{z}.$$ We thus work with a K\"ahler structure on $ \cal M$ and a line bundle on it with a connection whose curvature is $- i \omega$. The rest of the prequantization technique can be analogously constructed as given in \cite{Basu:2009dy}. This equips us with prequantized Hamiltonian functions $ \hat \theta ^{ \prime} = - \frac{2 i}{k+2} \tau \partial _z + \pi z$ and $ \hat \phi ^{\prime}= \frac{2i}{k+2} \partial _z$. It is important to note the shift of $k$ by the dual Coxeter number of $SU(2)$ to $k+2$ which originates from the non-trivial Polyakov-Wiegman factor \cite{Labastida:1989xp} for non-abelian compact gauge groups. In a more rigorous fashion its appearance is explained due to non-anomalous connection construction on the Hilbert bundle in \cite{Axelrod:1989xt}, which guarantees finally the quantum Hilbert space to be independent of the complex structure initially chosen for quantization.

We finally impose the quantization conditions on the polarized wavefunctions $ \psi(z)$ \footnote{the apparent operator ordering ambiguity is unphysical, costing only up to a phase in the wavefunction}:
$$ \mathrm{e} ^ { i (k+2) m \hat \theta ^{ \prime}} \mathrm{e} ^ { - i (k+2)n \hat \phi ^{ \prime}} \psi (z) = \psi (z).$$
This is solved by level $r=k+2$ theta functions:
$$ \vartheta _{j,r}(z, \tau) =  \sum_{n \in \mathbbm{Z}} \exp \left[2\pi i r \tau \left( n + \frac{j}{2r}\right)^2 + 2\pi i r z \left(  n + \frac{j}{2r}\right)\right]$$ with $ j = -r+1,\dots,r $ (since $ \vartheta _{j+2r , r} (z , \tau) =\vartheta _{j , r} (z , \tau)$). We will now construct the Weyl invariant subspace of this $ 2r $ dimensional vector space. Weyl invariance on $ \cal M$ means identifiction of $z$ with $-z$ \footnote{this is so because the traces of the holonomies \eqref{hom} are gauge invariant rather than $ h[\alpha] , h[\beta ]$ themselves and the traces do not distinguish between $( \theta, \phi ) $ and $ (-\theta, - \phi)$. This is another statement of Weyl invariance.}. Observing that $ \vartheta _{j , r} (-z , \tau) =\vartheta _{-j , r} (z , \tau)$ we project to the Weyl-odd subspace consisting of the $r-1 = k+1$ vectors:
$$ \vartheta ^{-}_{j,r}(z, \tau) =  \vartheta _{j , r} (z , \tau) -\vartheta _{-j , r} (z , \tau) ~~ j=1,\dots r-1 .$$
As per \cite{Axelrod:1989xt} one should now consider a `quantum bundle' over the space of complex structures $ \tau $ with fibres as the Hilbert space we have just found. The physical states should be parallel sections of this new bundle with respect to a projectively flat connection of the `quantum bundle'. Those vectors turn out to be:
\bea \label{vec}
 \psi _{ j , k} (z, \tau) = \frac{\vartheta ^{-}_{j+1,r}(z, \tau)}{\vartheta ^{-}_{1,2}(z, \tau)} ~~~~~~~~ j = 0 , \dots k
\eea
By taking the ratio of two Weyl-odd function we thus found the Weyl invariant vector space as desired. This space is orthonormal and serves as the required Hilbert space. 
\subsection{Gluing and $L(p,q)$}
We know that the mapping class group $SL(2, \mathbb{Z})$ or rather $SL(2, \mathbb{Z} )/\mathbb{Z} _2$ of $T^2$ is `generated' by two modular transformation elements $T, S$. Any general element $U$  of $SL(2, \mathbb{Z})$  can be expressed  as
$$ U = S \prod _{s=1}^{t-1} \left(T ^{m_s} S \right).$$
In its 2 dimensional representation $U$ produces $L(p,q)$ by gluing two solid tori for \cite{Freed:1991wd}
$$U = \begin{pmatrix}
q & b \\
p & d
\end{pmatrix}$$
The above representation of $U$ in terms of $T , S$  implies the following identity \cite{Jeffrey:1992tk}:
\bea \label{cont}
p/q = - m_{t-1} + \frac{1}{m_{t-2}-\dfrac{1}{\dots - \dfrac{1}{m_1}}}
\eea
The Chern-Simons-Witten invariant or the partition function is given by \cite{Witten:1988hf},
$$Z(r) _{L(p,q)} = \langle \psi _{0,k} |U| \psi _{0,k} \rangle$$
and it is independent of the parameters $b,d$ \cite{Jeffrey:1992tk}.
From the knowledge of action of $S$ and $T$ on theta functions we can evaluate these matrix elements. In the canonical 2-framing this was evaluated to be
\bea \label{Z1}
Z(r)_{L(p,q)} = -\frac{i}{\sqrt{2rp}}\exp \left( 6 \pi i s(q,p) /r\right) \sum _{\pm} \sum _{n=1}^{p} \exp \left( \frac{2\pi iq rn^2}{p} + \frac{2 \pi i n (q \pm 1)}{p} \pm \frac{\pi i}{rp}\right)
\eea
$$\mbox{where } s(q,p) = \sum _{l=1} ^{p-1} \dfrac{l}{p} \left( \dfrac{l q}{p }- \left[ \dfrac{lq}{p}\right] - \dfrac{1}{2}\right)$$ is the Dedekind sum defined in terms of the floor function [ ].
\subsection{Sum over topologies and finiteness of the partition function}
We note from the construction of $ \mathcal{H}_{T^2}$ \eqref{vec} that the dimension of the Hilbert space is $ r _{\SR} -1$ corresponding respectively to the '+' type and '-' type CS sectors. This is meaningful only when $ r _{\SR} -1  \in \mathbb{N}$ (excluding zero). These conditions come out to be stringent and restrict the parameters of the theory. Since $r _{\SR}-2 = k_{\SR} = \dfrac{l(1/\gamma \pm 1)}{8G}$, we have (when $\hbar$ and $c$ are restored suitably) \footnote{$l_p$ is the three dimensional Planck length $l _p = G \hbar /c^3$} the following restrictions
 \bea \label{param}
 a:= \dfrac{l}{8 l_p} =s/2 ~~~s\in \mathbb{N} ~~~~\mathrm{and}~~~~ \gamma = \dfrac{a}{(a-1)+t} ~~~t\in \mathbb{N}. \eea
These restrictions are the prototypes of any topological field theory \cite{Atiyah:1990dn}. One may be tempted to compare these with those appearing in \cite{Witten:2007kt} for $k_{\SR}$, where the unequal CS parameters are prescribed with discrete values in context of gravity. The apparent difference is due the choice of a different background topology used in \cite{Witten:2007kt}.

These nontrivial restrictions which validate the quantization (through positivity of the dimension of the Hilbert space) does not allow $ \gamma \rightarrow \infty $ which was again the starting point of the ordinary theory \eqref{action1}.
It is also interesting to see that the set of allowed value of $\gamma$ also includes 1, the `chiral' point for $t=1$. This motivates us strongly to study the corresponding Chiral limit of the underlying dual-CFT, if any. 

Leaving those issues for later discussion we now return to our original problem and express the gravity partition function (henceforth by gravity partition function we mean the partition function for the first order gravity )  as the product of the partition functions of `+' type and the `-' type theories \eqref{action3}:
\bea \label{partf} Z ^{ \mathrm {Grav}}_{L(p,q)} = Z(r_{\SP})_{L(p,q)} Z(r_{\SP})_{L(p,q)} \eea 
Full gravity partition function would on the other hand be stated after summing over all topologies ie 
$$Z^{ \mathrm{tot}} = \sum _{p=1}^{\infty} \sum _{ \substack{q( \mathrm{mod} p)\\ (q,p)=1}} Z ^{ \mathrm {Grav}}_{L(p,q)}$$
This final sum is where one encounters the divergence as explained in \cite{Castro:2011xb} through sums of kind $\sum _{ \substack{q( \mathrm{mod} p)\\ (q,p)=1}} 1= \phi (p)$, the Euler totient function. For the purpose of comparison with \cite{Castro:2011xb} and study the convergence property of our partition function we choose a particular classical saddle for which the sum over $n$ in \eqref{Z1} is replaced by a particular value of $ n = \frac{q \pm 1}{2}$ respectively for the `+' and the `-' type theory instead of taking the corresponding sum in \eqref{partf}. In order to bring in clarity further simplification is made through assuming $a$ to take only integral values and $a/\gamma \in 2 \mathbb{N}$. However these simplifications do not alter the final convergence properties of the sum. Using \eqref{Z1} in a more illuminating form \footnote{Let $A$ be the set of all such integers $q (\mathrm{mod} p)$ with $(q,p)=1$. It is easy to see that $\{q^* (\mathrm{mod} p)| q q^* =1( \mathrm{mod} p)\} =A$. This property has been used.} we have explicitly:
\bea \label{final}
Z^{ \mathrm{tot}} &=& -\frac{1}{2 \sqrt{ r_{\SP}r_{\SM}}}\sum _{p=1}^{\infty}\frac{1}{p} \sum _{ \substack{q( \mathrm{mod} p)\\ (q,p)=1}}\exp \left( 6 \pi i s(q,p) /R_{+}\right) \exp \left( \frac{\pi i}{p} \left( 2a + \left(q +q^*\right)\left( a/\gamma +2\right)\right)\right)\times \nonumber \\
&& \left[ \mathrm e ^{ \frac{\pi i }{p R_+} + \frac{2\pi i}{p} (q+1)}+\mathrm e ^{ -\frac{\pi i }{p R_+} + \frac{2\pi i}{p} (q-1)} -\mathrm e ^{ \frac{\pi i }{p R_-} + \frac{4\pi i}{p} }-\mathrm e ^{ \frac{-\pi i }{p R_-}}\right]
\\&& \hspace{-1.6cm} \mbox{where } \frac{1}{R _{\pm}} = \frac{1}{r _{\SP}} \pm \frac{1}{r _{\SM}} \nonumber
\eea
It is now easy to see that all the terms in the $q$ summand are $q$ dependent and the divergence producing totient function does not occur. However since no closed form of the $q$ sum is available, for the purpose of explicit checking we go to the limit where $ \gamma > 0$ is small $(\ll 1)$. Since the coupling constants become effectively large in this limit the partition function contains the expressions up to one loop. From \eqref{param} one observes that this limit is consistent with our quantization program by fixing $a$ and pushing the integer $t $ large. In this limit $ \dfrac{1}{R_+}\sim \dfrac{2 \gamma}{a}$ and $ \dfrac{1}{R_-}\sim \dfrac{2 \gamma ^2}{a} $ are both small. Out of the $ \gamma$ terms appearing as polynomials in the exponentials of \eqref{final} ie, $ \dfrac{1}{ \gamma}, 1, \gamma , \gamma ^2$ we keep  $ \dfrac{1}{ \gamma}, 1$ and neglect the last two. This implies
\bea \label{final2}
Z^{ \mathrm{tot}}=-\frac{\gamma}{a}\left(1-\frac{2\gamma}{a}\right)\sum _{p=1}^{\infty}\frac{1}{p} \mathrm e ^{ \frac{2\pi i a}{p}}\cos(2\pi /p)\Bigg[ S(\frac{a}{2\gamma} +2, \frac{a}{2\gamma} +1;p) - \mathrm e ^{ \frac{2\pi i }{p}}S(\frac{a}{2\gamma} +1, \frac{a}{2\gamma} +1;p)\Bigg]
\eea

$$ S( \alpha,\beta;p)=\sum _{ \substack{q( \mathrm{mod} p) \\ (q,p)=1}} \exp \left( 2\pi i ( \alpha q +\beta q^*)/p\right)$$
 Expanding the exponential and the cosine functions in the inverse power of $p$ we obtain an infinite series of Kloosterman zeta functions defined by
\bea L(m,n;s) = \sum _{p=1}^{\infty} p^{-2s} S(m,n;p). \nonumber\eea Kloosterman zeta function is again analytic in the region $\Re s > 1/2$.

Now, as we are in the {\it small} $\gamma $ regime, the summand in \eqref{final2} can well be approximated as
\bea \label{anys}
&&\sum _{p=1}^{\infty}\frac{1}{p} \mathrm e ^{ \frac{2\pi i a}{p}}\cos(2\pi /p)\left( 1- \mathrm e ^{ \frac{2\pi i }{p}}\right)S(\frac{a}{2\gamma} , \frac{a}{2\gamma};p)\nonumber\\
&=& \sum _{m,n,r=0}^{\infty} \frac{(2\pi i)^{r+n+2m+1}}{r+1} \frac{a^n}{(2m)! n! r!}\sum _{p=1}^{\infty} p^{-(r+n+2m+2)} S(\frac{a}{2\gamma} , \frac{a}{2\gamma};p)\nonumber \\
&=&  \sum _{m,n,r=0}^{\infty} \frac{(2\pi i)^{r+n+2m+1}}{r+1} \frac{a^n}{(2m)! n! r!} L(\frac{a}{2\gamma} , \frac{a}{2\gamma};\frac{r+n+2m}{2}+1)
 \eea
The good news is that we get a series of $L(\frac{a}{2\gamma} , \frac{a}{2\gamma};s)$ with $s \geq 1$. Hence the partition function is free from divergences. Had we set $ a/\gamma +2 = 0$, the second Kloosterman sum would have reduced to the totient function. That is a potential source of singularity, which is obvious since its zeta function is expressed in terms Riemann zeta function and $ \zeta (1)$ is singular. We again see that the finiteness of the parameter $ \gamma$ saves us from having a meaningless quantization.

Here we wish to point out that we are evaluating the partition function in the case of small $\gamma$. This again corresponds to large CS couplings $k_{(\SR)}$. However quantum CS theory dictates that large coupling means first quantum correction \cite{Jeffrey:1992tk}. In that sense \eqref{final2} or \eqref{anys} corresponds to one loop result.
\section{The metric counterpart and the TMG story}
The key relation connecting the first order formalism and metric regime is : $\langle e_{\mu}, e_{\nu}\rangle = g_{\mu \nu}$. It  should be supplemented  with the torsionless condition ensuring the geometry to be Riemannian. If one starts with the action \eqref{action3}, one gets this condition \eqref{eomb} as an equation of motion. Solving this equation makes \eqref{action2} the well known gravitational Chern Simons and \eqref{action1} the Einstein Hilbert action provided we use only the invertible subset of vierbeins from \eqref{eomb} . The action \eqref{action3} becomes TMG with $\gamma $ playing the role of topological mass. It is not surprising that dynamics of TMG and that of \eqref{action3} are quite different; including equations of motion and canonical structures. The most important feature perhaps is that TMG has local degree of freedom which is absent in the theory described by \eqref{action3} and one should not expect similarity in their quantum theories. However TMG being the closest kin to our theory in metric version, for a completion we present a comparative study with quantum TMG focussing its convergence properties as worked out in detail in \cite{Castro:2011ke}.

To be more precise, we first focus on what is meant by quantum dS TMG. This issue, as we have already mentioned, has been exhaustively studied in \cite{Castro:2011ke}.8u. The one loop partition function is showed there to converge. Denoting by E, the contributions coming from pure Einstein Hilbert theory with cosmological constant and by MG, the ones coming from massive graviton modes, they show that:
\bea
\sum _{p=1}^{\infty} \sum _{ \substack{q( \mathrm{mod} p)\\ (q,p)=1}} Z^{(0)}_{\mathrm E} Z^{(0)}_{\mathrm MG} Z^{(1)}_{\mathrm E} \sim 
\sum _{r=0}^{\infty} \frac{(2\pi a) ^r}{r!} L(\frac{a}{2\gamma},\frac{a}{2\gamma};\frac{r}{2}+\frac{1}{2}) + \mbox{trivially analytic terms}.
\eea
One can now compare this with \eqref{anys}. The interesting fact is that here the term corresponding to $r=0$ in the sum of the RHS is the source of divergence since it corresponds to Kloosterman zeta function with $s=1/2$. But it is also showed in \cite{Castro:2011ke} that when one includes $Z^{(1)}_{\mathrm MG}$ as the product and then performs the sum over $p$, the divergence is eaten up. This means that up to one loop calculation they have
$$Z =\sum _{p=1}^{\infty} \sum _{ \substack{q( \mathrm{mod} p)\\ (q,p)=1}} Z^{(0)}_{\mathrm E} Z^{(0)}_{\mathrm MG} Z^{(1)}_{\mathrm E} Z^{(1)}_{\mathrm{MG}}.$$
The expression of $Z^{(1)}_{\mathrm{MG}}$ as given in \cite{Castro:2011ke} is far too complicated for the above expression to be analytically simplified and compared with \eqref{anys}. But the mechanism through which the divergence in the above expression is controlled by $Z^{(1)}_{\mathrm MG}$ is very similar to the way we showed \eqref{anys} to be finite. In essence both our topological theory of gravity and TMG (dynamical) have finite and {\it{similarly}} convergent partition functions. Since these theories are classically different this fact seems to be quite surprising. That TMG is derived as a metric version of our theory may however qualitatively explain this similarity in  partition functions up to one loop. We conclude that although the finiteness of TMG could be ascribed to its propagating graviton modes , our theory \eqref{action3}, being devoid of massive gravitons still yield a reasonably similar convergent partition function.
\section{Conclusion}
The take home message of our analysis can be summarized as follows :
\begin{enumerate}
\item  Construction of the associated Hilbert spaces on the torus surfaces is correct only for finite $\gamma$. These constructions spell out the set of allowed values of $\gamma$ and this does not include $ \gamma \rightarrow \infty$.
\item That finite values of $ \gamma $  can make the partition function divergence free is shown explicitly for $ \gamma \ll 1$. This is most important from point of view of the quantization of lens space gravity. 
\end{enumerate}

The fact that pure Einstein gravity has divergent partition function even at one loop and TMG is finite may seem to be a lucrative point of discussion in context of the work we present here. One can pass over to TMG (essentially dynamical) from action \eqref{action3}, which is topological, by imposing the torsionless condition. Hence they share the same parameter content. In the AdS sector however, this similarity is more pronounced as they have same dual CFTs. Whereas in present case, such an analogy is premature, since dual CFT in 3D de Sitter gravity is yet to be understood. Any progress in this front would surely shade light on the proposed dS/CFT \cite{Strominger:2001pn} correspondence (which works in 4 dimensions) in 3 dimensions and on its gravitational interpretation.


On the other hand, the finiteness brought in by the gravitational Chern Simons term of TMG also may be interpreted in light of \eqref{action3}. This being parity odd, there are phases in the partition function. Control of the divergence can be ascribed to this fact. This explanation works in the perturbative regime for TMG at least, as shown in \cite{Castro:2011ke}. 
Our result being finite is in conformity with TMG.

Another point of interest which we leave for future study is the interpretation of the theory when $ \gamma \rightarrow 1$. In the AdS paradigm an analogous point in parameter space has been shown to have critical CFT dual \cite{Li:2008dq, Maloney:2009ck}. In light of the proposed dS/CFT \cite{Strominger:2001pn} framework this may serve as an exciting evidence for dual critical CFT. 
\section*{Acknowledgments}
 RB thanks Council for Scientific and Industrial Research (CSIR), India, for support through the
SPM Fellowship SPM-07/575(0061)/2009-EMR-I. The authors also thank the anonymous referee for comments which helped improving the manuscript considerably.

\end{document}